# Process intensification strategies for enhanced holocellulose solubilization: Beneficiation of pineapple peel waste for cleaner butanol production

Manisha A. Khedkar[1], Pranhita R. Nimbalkar[1], Sanjay P. Kamble[2], Shashank G. Gaikwad[2], Prakash V. Chavan[1,**], Sandip B. Bankar[3,*]

**\*Corresponding authors**:
\*Sandip B. Bankar, Department of Bioproducts and Biosystems, School of Chemical Engineering, Aalto University, P.O. Box 16100, FI-00076 Aalto, FINLAND.
\*\*Prakash V. Chavan, Department of Chemical Engineering, Bharati Vidyapeeth (Deemed to be University), College of Engineering, Dhankawadi, Pune-Satara Road, Pune - 411 043, INDIA.

E-mail addresses: pvchavan@bvucoep.edu.in (P.V. Chavan), sandipbankar@gmail.com, sandip.bankar@aalto.fi (S.B. Bankar).

## Abstract

Biorefinery sector has become a serious dispute for cleaner and sustainable development in recent years. In the present study, pretreatment of pineapple peel waste was carried out in high pressure reactor using various pretreatment-enhancers. The type and concentration effect of each enhancer on hemicellulose solubilization was systematically investigated. The binary acid (phenol + sulfuric acid) at 180 °C was found to be superior amongst other studied enhancers, giving 81.17% (w/v) hemicellulose solubilization in liquid-fraction under optimized conditions. Solid residue thus obtained was subjected to enzymatic hydrolysis that resulted into 24.50% (w/v) cellulose breakdown. Treated solid residue was further characterized by scanning electron microscopy and Fourier transform infrared spectroscopy to elucidate structural changes. The pooled fractions (acid treated and enzymatically hydrolyzed) were fermented using *Clostridium acetobutylicum* NRRL B 527 which resulted in butanol production of 5.18 g/L with yield of 0.13 g butanol/g sugar consumed. Therefore, pretreatment of pineapple peel waste evaluated in this study can be considered as milestone in utilization of low cost feedstock, for bioenergy production.

# 1. Introduction

The energy consumption in India is expected to grow 3.8% a year by 2035, owing to continual national progress for improved living, faster than all major economies in the world (Deb and Appleby, 2015). However, limited crude oil reservoirs with fluctuating market prices because of international politics will result into formation of energy gap (Zhao et al., 2015) that need to be addressed based on local resources. One of the possible solutions to overcome energy insecurity is to develop advanced fuels from sustainable and renewable resources (Bankar et al., 2013). Apart from many agricultural and related industries; fruit industry waste can be considered as an interested and viable feedstock (renewable resource) for production of liquid fuels. Pineapple peel (generated after juice processing) is reflected as low-cost feedstock thus suitable for production of value added biochemicals. Generally, pineapple peel biomass constitutes main structural carbohydrates such as cellulose, hemicellulose, and lignin, compactly stacked together to form a rigid and recalcitrant structure (Choonut et al., 2014). An efficient pretreatment can conquer the recalcitrant nature and provide cellulose accessibility during feedstock processing. Physical, chemical, physico-chemical, thermo-chemical, and biological processes are being used for biomass conversion over the years (Adsul et al., 2005; Liang et al., 2017; Madadi et al., 2017). Incidentally, pretreatment of feedstock using steam explosion is increasingly gaining popularity in biorefinery sector owing to its favorable environmental and techno-economical aspects (Pocan et al., 2018; Scholl et al., 2015; Zhang et al., 2017). Steam explosion process disrupts the rigid structure of biomass to freed cellulose which can further hydrolyzed to yield monomeric sugars.

Voluminous literature is available on pretreatment techniques using acids (sulfuric, hydrochloric, nitric, phosphoric, maleic, and oxalic acids along with ferric chloride and zinc chloride) and alkalis (sodium hydroxide, potassium hydroxide, and ammonium hydroxide) (Ayeni and Daramola, 2017; Kootstra et al., 2009; Tekin et al., 2015; Adsul et al., 2005; Kumar and Murthy, 2011; Silva et al., 2017; Yamamoto et al., 2014). Researchers have used dilute acid pretreatments particularly in steam explosion reactor for breakdown of complex lignocellulosic materials (Gonzales et al., 2016; McIntosh et al., 2016). Dagnino et al. (2013) studied dilute acid pretreatment with due emphasis on operational parameters such as temperature, residence time and different concentrations of acids used. They have reported positive influence on xylose release and inhibitor generation. Besides, our previous study also revealed that the dilute sulfuric acid pretreatment of pineapple peel waste carried out in an autoclave at 121 °C resulted in total sugar release of 97 g/L (Khedkar et al., 2017a). Furthermore, it will be advantageous to explore the concept of binary acid for effective hemicellulose solubilization which is expected to lower the inhibitor formation (Pal et al., 2016a).

In this article, the use of pretreatment-enhancers *viz.* single acids, binary acid, and mixed acid were explored to see their influence on structure, morphology, and enzymatic treatment of biomass. The effect of each enhancer namely hydrogen peroxide, phenol, oxalic acid, and sulfuric acid was investigated individually and/or in combination with respect to maximum sugar release. Interestingly, hydrogen peroxide is simple peroxide that is used for oxidative delignification in order to supply oxygen into the pretreatment process (Ayeni and Daramola, 2017). Phenol is a carbolic acid, an aromatic organic compound, weakly acidic in nature and acts as free radical scavenger in aqueous solution (Lamminpää et al., 2015). Oxalic acid is a biomimetic; possesses better selectivity for hemicellulose by structurally mimicking catalytic functional groups (Pal et al., 2016a). In consideration of abovementioned characteristics of pretreatment-enhancers, it is interesting to evaluate their synergistic effect during pretreatment and subsequently on enzymatic hydrolysis in order to improve the sugar release. The binary acid chosen for this study were a combination of mineral acids such as hydrogen peroxide, phenol, and oxalic acid. Sulfuric acid (common mineral acid in all binary acid study) was used as a primary enhancer in current study because of high acidity, low cost, short residence time, and process simplicity (Pal et al., 2016b).

The objective of present study was to evaluate the influence of single and binary acid pretreatment-enhancers on hemicellulose hydrolysis. Moreover, the hydrolysis performance was estimated in terms of sugar release and inhibitor generation. Subsequently, steam exploded solid residue was enzymatically hydrolyzed to obtain simple sugars. Furthermore, fermentability of hydrolysate was tested using Clostridium acetobutylicum NRRL B 527 to check the acetone-butanol-ethanol (ABE) production. The structural and functional group analysis of pre-treated samples were also carried out using scanning electron microscopy (SEM) and fourier transform infrared spectroscopy (FTIR) to examine alterations during pretreatment.

## 2. Materials and methods

2.1. Materials

Dextrose, dipotassium hydrogen phosphate, potassium dihydrogen phosphate, ammonium acetate, biotin, thiamin, *p*-aminobenzoic acid, peptone, L-cysteine hydrochloride, sodium chloride, magnesium sulfate, manganese sulfate, iron sulfate, sodium hydroxide, butyric acid, acetic acid, sulfuric acid, phenol, butanol, acetone, and ethanol were purchased from SRL Ltd, Mumbai, India. Sodium acetate and hydrogen peroxide were procured from Sigma Aldrich, Bangalore, India. Folin-Ciocalteu reagent, oxalic acid, sodium carbonate, meat extract, maltose, and yeast extract were obtained from Himedia, Mumbai, India.

2.2. Pretreatment of pineapple peel

Pineapple peel waste was obtained from local market of Pune, Maharashtra, India. Sample preparation and proximate analysis was done as reported in previous study (Khedkar et al., 2017a).

*2.2.1. Screening of pretreatment-enhancers*

The pineapple peel biomass was subjected to pretreatment experiments with inclusion of single enhancers and their combinations, in order to study the effect on hemicellulose solubilization, cellulose hydrolysis, and lignin removal. The percent concentration used for each enhancer (sulfuric acid, phenol, and hydrogen peroxide) was on v/v basis except for oxalic acid (w/v). Control experiment (phenol + sulfuric acid without sample) was performed to calculate the actual phenolics produced during pretreatment. A laboratory autoclave (Bio-Technic BTI 02, India) was used during all treatments at 121 °C for 15 min with biomass to liquid ratio of 1:5 (20 g dry biomass to 100 mL liquid mixture). Initially, hydrogen peroxide, phenol, and oxalic acid were tested individually at 1%. Sulfuric acid was used at its optimum concentration of 1.3% as used in previous study (Khedkar et al., 2017a). Subsequently, the combinations of two acids and mixed acids were also tested in the concentration range of 0.5-2% (Table 1). Reaction mixtures were separated into solid and liquid fractions after each treatment. The residual solid was dried in a conventional oven at 100 °C till constant weight was achieved, for its further use in enzymatic hydrolysis. Liquid fraction was stored in a refrigerator (4 °C) for sugar and inhibitor analysis.

2.2.2. Steam explosion

Steam explosion experiments were performed in 2 L high pressure SS-316 (100 kg/cm$^2$) reactor (make- Amar equipment Pvt. Ltd., Mumbai, India) equipped with a temperature sensor, pressure transducer, pressure gauge, vent valve, safety rupture disc, gas inlet-valve, liquid sampling valve, agitator, pitched blade impeller motor, solenoid valve, and a sampling port. Reaction pressure, temperature, and agitator speed were continuously monitored during the experiments using SCADA control panel. The experiments were carried out at 150, 180, and 200 °C with incorporation of binary acid at concentration between 1 and 1.5%. Sulfuric acid (1.3%) was used as a control to compare the performance with respect to sugar release (Table 2). The 40 g sample (dry) was added into 200 mL dilute acid to carry out reaction at 200 rpm for 10 min. The hydrolysates were filtered using Whatman filter paper (pore size 20-25 mm) for solid-liquid separation. The liquid fraction was stored at 4 °C and solid residue was dried at 100 °C for their subsequent use. All the experiments were carried out in triplicate and results presented are with average value.

2.3. Characterization of pretreated solid-fraction
*2.3.1. Compositional and morphological analysis*
The cellulose, hemicellulose, lignin, moisture, and ash content were determined as per previous protocol (Khedkar et al., 2017a). The morphology of untreated and pretreated samples were investigated using SEM (VEGA 3SBU, TESCAN, Bron, Czech Republic). Samples were adhered to carbon tape and sputter-coated with gold. The images were captured at an acceleration voltage of 10 kV.

*2.3.2. FTIR spectroscopic analysis*
The untreated and treated samples were evaluated for their surface chemistry using FTIR spectroscopy from Bruker in the range of 500-4000 (1/cm). Data obtained was analyzed using Opus software. The crystallinity was determined by three methods *viz.* total crystallinity index (TCI), lateral order index (LOI), and hydrogen bond intensity (HBI) (Ayeni and Daramola, 2017; Ciolacu et al., 2011; Poletto et al., 2014). TCI (correspond to the C-H stretching) was determined by ratio of wavenumber (1/cm) at 1378/2900. The ratio of 1437/899 was used to estimate LOI (correspond to a $CH_2$ bending vibrations), and 3400/1320 was used to estimate HBI.

2.4. Enzymatic hydrolysis of pretreated solid-residue
The acid and enzymatic treatments were carried out separately. Solid fractions received after steam explosion treatment was neutralized to pH 7 to remove excess acid present. The samples were dried at 100 °C till it reached constant weight. Dried residues were then hydrolyzed by cellulase at fixed unit (Cellic CTec2, SAE 0020 - Cellulase, enzyme blend). The optimized conditions for enzymatic hydrolysis were taken from a study reported by Harde et al. (2016). Enzyme (60 FPU/g substrate) was added in 10 g pretreated solid residue together with 100 mL of sodium acetate buffer (pH - 5.5, 100 mM). The reaction mixture was kept at 50 °C and 120 rpm for 72 h. After 72 h, the reaction mixtures were boiled for 1 min to deactivate the enzyme. The mixture was centrifuged to remove insoluble solids and supernatant was analyzed by high performance liquid chromatography (HPLC) for sugar estimation.

2.5. Detoxification of pretreated liquid-fraction
Liquid fraction obtained after steam explosion contained many fermentation inhibitors such as acetic acid, total furans, and total phenolics. Hence, it was thought desirable to detoxify the liquid fraction as per process detailed in previous study (Khedkar et al., 2017a). Solid-fractions after steam explosion were also enzymatically hydrolyzed to produce fermentable sugars. The resulted liquid stream did not contain any fermentation inhibitors and hence detoxification step was excluded.

2.6. ABE fermentation
*2.6.1. Microorganism*
The microbial strain *C. acetobutylicum* NRRL B-527 was generously gifted by ARS (Agriculture Research Services) culture collection, USA. The details of revival procedure and preparation of spore suspension was described in previous work (Nimbalkar et al., 2018). Spore suspension was activated by heat shock treatment and activated cells (2% v/v) were grown into 100 mL sterile reinforced clostridial medium for 18-20 h. 5% (v/v) actively growing cells (cell turbidity (OD560nm) - 1.09) were used as an inoculum during fermentation.

*2.6.2. Mixed sugar (pentose and hexose) fermentation*
The fermentation experiments were performed in 100 mL air tight bottles with 80 mL working volume. Standard P2 medium reported by Bankar et al. (2012) was used as control for ABE fermentation. Liquid stream obtained from enzymatic treatment and liquid fraction after steam explosion were pooled together for fermentation experiments. The liquor thus obtained was diluted to maintain total sugar concentration to be 60 g/L (glucose 38.69 g/L, fructose 8.2 g/L, xylose 14.29 g/L, and arabinose 0.88 g/L). Subsequently, other fermentation medium components were supplemented to the processed liquor and pH was adjusted to 6.5 (Nimbalkar et al., 2017). The fermentation medium was purged with nitrogen to maintain anaerobic conditions and autoclaved at 121 °C for 20 min. Fermentation was started by inoculating 5% (v/v) 20 h old inoculum (OD560nm 1.09) and incubated at 37 °C for 120 h. The samples were withdrawn at 24 h time interval and analyzed for total solvents (butanol, acetone, and ethanol) and total acids (acetic acid and butyric acid) using gas chromatography. All the experiments were done in triplicate and results are presented as average values ± standard deviation.

2.7. Analytical method
*2.7.1. Sugars and inhibitors analysis of pretreated sample*
The individual sugars (glucose, xylose, mannose, and arabinose) were analyzed by HPLC system (Dionex India Ltd.) equipped with a refractive index detector and an ion exclusion column (Biorad Aminex, HPX 87-H). Maltose was used as an internal standard during analysis. Mobile phase used was 80 mM $H_2SO_4$ at a flow rate of 0.6 mL/min with operating temperature of 50 °C and sample volume of 100 mL. The phenol sulfuric acid method was also used to estimate total sugar concentration (Dubois et al., 1956). Total phenolic content was estimated using Folin-Ciocalteu method (Maurya and Singh, 2010). Besides, the total furans (Furfural and 5-hydroxymethyl furfural) were estimated using a spectroscopic method based on difference in the absorbance values taken at 284 and 320 nm (Martinez et al., 2000).

*2.7.2. Solvents and acids analysis*
The solvents (acetone-butanol-ethanol) and acids (acetic and butyric) estimation were done using gas chromatography with automatic headspace sampler (Agilent Technologies 7890B) equipped with a flame ionization detector and AB-INNOWAX capillary column (30 m×0.32 mm× 1 mm). Temperatures were maintained at 200 and 250 °C for injector and detector, respectively with sample volume of 0.5 mL.

**3. Results and discussion**
3.1. Pretreatment of pineapple peel
The proximate analysis of pineapple peel was performed as detailed in earlier study (Khedkar et al., 2017a).
*3.1.1. Screening of pretreatment enhancers*
Effect of single acid, binary acid, and mixed acid were investigated at 121 °C under specified conditions intended for hemicellulose solubilization and indirectly on cellulose hydrolysis (with the help of enzymes). Our previous study revealed that dilute sulfuric acid treatment above 2% was ineffective in regard to sugar release from pineapple peel (Khedkar et al., 2017a). Therefore, it was decided to perform initial screening of pretreatment enhancers (hydrogen peroxide, phenol, sulfuric acid and oxalic acid) at appropriate concentrations. The monosaccharide concentration, fermentation inhibitors, and pH of the hydrolysates are as shown in Table 1. Dilute sulfuric acid treated samples showed higher total sugar release to be 118.06 g/L, which is higher than other individual treatments (Table 1).

Table 1 Screening of single acid catalyst and its combinations of binary acid pretreatment (121 °C, 15 min, 20 % solid loading) of pineapple peel, solid and liquid fractions analysis

| R.N. | Particulars | Hydrolysate fraction analysis (g/L) | | | | Solid fraction analysis (g/L)* | |
|---|---|---|---|---|---|---|---|
| | | Total furans$ | Total phenolic | Acetic acid | pH | Total sugar# | Hexose sugar** | AIL |
| **Single acid pretreatment-enhancer** | | | | | | | | |
| 1 | 1.3 % Sulfuric acid | 4.7 | 5.5 | 5.2 | 1.9 | 118.06 | 97.69 | 3.8 |
| 2 | 1 % Oxalic acid | 7.3 | 4.72 | 4.99 | 3.1 | 97.19 | 99.66 | 3.3 |
| 3 | 1 % Hydrogen peroxide | 0.91 | 7.07 | 1.14 | 3.7 | 82.38 | 92.25 | 3.7 |
| 4 | 1 % Phenol | 0.98 | 14.52 | 3.36 | 4.6 | 95.58 | 113.28 | 1.9 |
| **Binary acid pretreatment-enhancer** | | | | | | | | |
| Oxalic acid + sulfuric acid (%) | | | | | | | | |
| 5 | 0.5+1.3 | 5.42 | 8.25 | 3.28 | 2.2 | 126.6 | 77.65 | 3.28 |
| 6 | 1+1.3 | 6.71 | 7.18 | 3.28 | 2.0 | 126.6 | 78.17 | 3.28 |
| 7 | 2+1.3 | 4.79 | 6.89 | 3.48 | 2.1 | 126.7 | 74.58 | 3.48 |
| Hydrogen peroxide + sulfuric acid (%) | | | | | | | | |
| 8 | 0.5+1.3 | 1.15 | 6.16 | 5.11 | 2.4 | 117.6 | 76.21 | 4.5 |
| 9 | 1+1.3 | 2.24 | 8.06 | 5.12 | 2.4 | 118.9 | 78.77 | 4.1 |
| 10 | 2+1.3 | 2.63 | 9.69 | 5.91 | 2.1 | 119.8 | 77.01 | 4.1 |
| Phenol + sulfuric acid (%) | | | | | | | | |
| 11 | 0.5+1.3 | 4.31 | 11.5 | 5.72 | 2.1 | 154.6 | 75.23 | 3.6 |
| 12 | 1+1.3 | 3.79 | 15.14 | 5.39 | 2.1 | 168.3 | 94.12 | 3.5 |
| 13 | 2+1.3 | 4.05 | 14.41 | 5.39 | 2.1 | 159.2 | 80.12 | 3.0 |

AIL: Acid insoluble lignin of treated solid residue
$ indicate the total furans in mg/L
# Designate the total sugar of acid hydrolysates with contribution of xylose, glucose, arabinose, and fructose
* Indicate solid residue obtained after treatment and used for enzymatic hydrolysis
** Designate the glucose sugar obtain after enzymatic hydrolysis

Oxalic acid, phenol, and hydrogen peroxide resulted in a total sugar release (g/L) of 97.19, 95.58, and 82.38, respectively and individual sugars can be seen in Supplementary Table S1. Interestingly, dilute sulfuric acid treatment provided 59.92% total sugar yield which is around 10e15% higher than other individual enhancers tested. This might be due to the influence of lower pH of used acid. Sulfuric acid treated hydrolysate showed relatively lower pH of 1.66 when compared with others such as oxalic acid (3.15), hydrogen peroxide (3.7), and phenol (4.6). Moreover, Pal et al., (2016a) also stated positive influence of low pH during hydrolysis which in turn enhances hemicellulose solubilization with reduced time of operation. However, extremely low pH may corrode the reactors that affect process efficiency and thus may increase the capital investment. This can be tackled by incorporating different combinations of pretreatment-enhancers which may possibly help to reduce the pretreatment cost with higher process efficiency.

The fermentation inhibitor formation has been well recognized during pretreatment processes (Amiri and Karimi, 2015; Dussan et al., 2014). Therefore, fermentation inhibitors namely acetic acid, total furan, and total phenolics from pretreated feedstock were analyzed (Table 1). As a result, phenol and hydrogen peroxide treatment showed lower levels of acetic acid (1.14 and 3.36 g/L) and total furans (0.91 and 0.98 g/L). The total phenolics measured in the experiment had some contribution (~50%) from phenol-treatment during enhancer study. Hence, control experiments without sample were performed to calculate the phenolics. The phenol treated samples thus showed higher total phenolics (14.52 g/L) than hydrogen peroxide (7.02 g/L) and other samples (Table 1). Additionally, sulfuric acid treatment also leads to alteration in lignin structure or lignin distribution in sample during the treatment. Incidentally, it was examined by FTIR and SEM analysis (for details refer section 3.2), which confirmed the changes in lignin orientation. Furthermore, compositional analysis of solid-fraction pointed out presence of lower level of total lignin which clearly shows that lignin was removed in liquid-fraction of the process (unpublished data).

Although, better yield of xylose sugars was obtained through sulfuric acid treatment, the formation of inhibitors such as acetic acid (5.2 g/L), total furans (4.7 mg/L), total phenolics (5.5 g/L), and acid insoluble lignin (3.8 g/L) were also found to be higher compared to other treatments (Table 1). Therefore, in order to maximize sugar release along with minimum inhibitor generation, a systematic study pertaining to combined effect of different acids was thought to be desirable. Based on aforesaid results, dilute sulfuric acid (1.3%) was selected as primary enhancer and kept common in binary acid study whereas other enhancer concentrations were varied in the range of 0.5-2%.

Table 1 shows the hydrolysate compositions after pretreatment using different concentration of oxalic acid, hydrogen peroxide, and phenol in combination with sulfuric acid. The treatment by 1% oxalic acid + sulfuric acid showed low acetic acid levels. Additionally, 1% hydrogen peroxide + sulfuric acid resulted in lower total furans with slight increase in xylose release (around 5%) by catalyzing the decomposition and defibrillation of biomass structure. Furthermore, the varied concentrations of oxalic acid and hydrogen peroxide in combination with sulfuric acid were ineffective in context of xylose release (Table 1). Besides, 1% phenol + sulfuric acid (Table 1; run no. 12) led to lower inhibitors (5.39 g/L acetic acid, 3.7 mg/L total furans, and 15.14 g/L total phenolic) with higher total sugar release (85.44%). This may be because of synergistic action of binary acid that in turn resulted into effective breakdown of lignocellulosic structure. These results are in good agreement with Pal et al. (2016a) for sugarcane bagasse treatments wherein inhibitors concentration using binary acid (sulfuric acid with oxalic acid) was decreased as compared to individual sulfuric acid treatment.

Additionally, 0.5% phenol + sulfuric acid resulted in low xylose release (around 5%) which may be attributed to partial hydrolysis of hemicellulose. On the other hand, sulfuric acid + phenol (beyond 1%) showed reduced total sugar release (Table 1). Ayeni and Daramola (2017) carried out pretreatment of corn cob using binary acid (hydrogen peroxide/sodium hydroxide) and reported highest solubilization of cellulose (59%) and hemicellulose (79%). Wei et al. (2011) also conducted hot water pretreatment of filter paper strips using individual dilute sulfuric acid and in combination with ferric chloride and reported glucose enhancement of 28% in combination treatment.

Furthermore, the effect of mixed acid (sulfuric acid, oxalic acid, hydrogen peroxide, and phenol) was also evaluated at concentration of 0.5 and 1% in order to improve the sugar release from pineapple peel waste. Surprisingly, hemicellulose solubilization was found to be inferior; resulting in total sugar release of 146.2 g/L and 122.07 g/L at concentration of 0.5 and 1%, respectively. Overall, binary acid was found to be efficient at lower pretreatment temperature for partial solubilization of lignocellulosic components. Hence, binary acid was further selected to operate at higher reaction temperature to achieve near-to-complete sugar release. Corredor et al. (2008) reported the effectiveness of higher reaction temperature for soya bean hulls to rupture the hydrogen bonds with elevated pentose yield of 96%.

### 3.1.2. Steam explosion

Steam explosion instigates the physical and chemical changes in biomass which will help to enhance the hemicellulose solubilization and cellulose accessibility. Steam explosion is routinely carried out at relatively high temperature (150-200 °C) under trivial acidic environment (Sui and Chen, 2016). Moreover, elevated temperature accelerates the mobility of hydrogen ions and augments the hydrolysis for monomeric sugar release. Further, higher temperature under acidic condition is effective in triggering a series of hydrothermal reactions during which sugars also get degraded into toxic by-products (furan derivatives, phenolics, and acetic acid) (Timung et al., 2015; Farmanbordar et al., 2018). As expected, the increase in pretreatment temperature exhibited substantial improvement in total sugar release accompanied with increased inhibitor formation. For example, the treatment at 121 °C using dilute sulfuric acid (1.3%) resulted into total sugar release of 118.06 g/L (Table 2), whereas increment in temperature to 150 and 180 °C showed improvement in total sugar release to be 204.55, and 237.24 g/L, respectively (Table 2). Besides, similar total sugar release pattern was observed for binary acid treatment (oxalic acid and/or hydrogen peroxide + sulfuric acid) at 150 °C in comparison with sulfuric acid alone (Table 2). Phenol + sulfuric acid treatment at 150 °C also showed increased total sugar release up to 225.55 g/L (Table 2).

**Table 2** Effect of binary acid in steam explosion (at 150, 180, and 200 °C, 20 % solid loading, for 10 min) pretreatment using pineapple peel, analysis of solid and liquid fractions

| R.N. | Particulars | Hydrolysate fraction analysis (g/L) | | | | | Solid fraction analysis (g/L)* | |
|---|---|---|---|---|---|---|---|---|
| | | Total furans$ | Total phenolics | Acetic acids | pH | Total sugar# | Hexose sugar** | AIL |
| **Pretreatment at 150 °C** | | | | | | | | |
| 1 | Sulfuric acid (1.3 %) | 10.3 | 6.81 | 5.9 | 1.5 | 204.66 | 86.93 | 4.88 |
| 2 | Oxalic acids (1 %) + sulfuric acid (1.3 %) | 6.1 | 8.81 | 6.1 | 2.1 | 210.60 | 89.2 | 3.13 |
| 3 | Hydrogen peroxide (1%)+ sulfuric acid (1.3 %) | 6.9 | 7.19 | 4.4 | 2.1 | 208.98 | 90.07 | 4.31 |
| 4 | Phenol (1 %) + sulfuric acid (1.3 %) | 5.6 | 12.45 | 5.3 | 2.1 | 225.55 | 91.68 | 2.88 |
| **Pretreatment at 180 °C** | | | | | | | | |
| 5 | Sulfuric acid (1.3 %) | 14.7 | 8.7 | 4.0 | 1.5 | 237.24 | 80.44 | 4.9 |
| 6 | Oxalic acids (1 %) + sulfuric acid (1.3 %) | 8.15 | 8.19 | 6.1 | 2.1 | 245.32 | 81.44 | 3.08 |
| 7 | Hydrogen peroxide (1%)+ sulfuric acid (1.3%) | 10.5 | 9.68 | 5.3 | 2.0 | 240.21 | 80.33 | 4.68 |
| 8 | Phenol (1 %) + sulfuric acid (1.3 %) | 6.35 | 11.06 | 5.9 | 2.0 | 357.59 | 85.77 | 2.96 |
| **Pretreatment at 200 °C** | | | | | | | | |
| 9 | Sulfuric acid (1.3 %) | 5.24 | 4.06 | 7.91 | 1.7 | 111.9 | 70.24 | 6.57 |
| 10 | Oxalic acids (1%) + sulfuric acid (1.3%) | 4.75 | 3.88 | 7.87 | 1.7 | 101.52 | 71.11 | 4.04 |
| 11 | Hydrogen peroxide (1%)+sulfuric acid (1.3%) | 6.08 | 6.63 | 7.2 | 1.7 | 135.53 | 69.54 | 3.75 |
| 12 | Phenol (1 %) + sulfuric acid (1.3 %) | 4.16 | 3.4 | 7.5 | 1.9 | 279.87 | 75.67 | 3.68 |

AIL : Acid insoluble lignin of treated solid residue
$ indicate the total furans in mg/L
# Designate the total sugar of acid hydrolysates with contribution of xylose, glucose, arabinose, and fructose
* Indicate solid residue obtained after acid treatment and used for enzymatic hydrolysis
** Designate the glucose sugar obtain after enzymatic hydrolysis

Likewise, combination of phenol + sulfuric acid treatment at 180 °C solubilized the highest hemicellulose (81.17%) including xylose (92.48 g/L), fructose (60 g/L), and arabinose (6.44 g/L) in the liquid fraction (Fig. 1). As the reaction temperature increased; amorphous cellulose gets hydrolyzed along with hemicellulose in liquid-fraction (198.56 g/L glucose) as can be seen in Fig. 1.

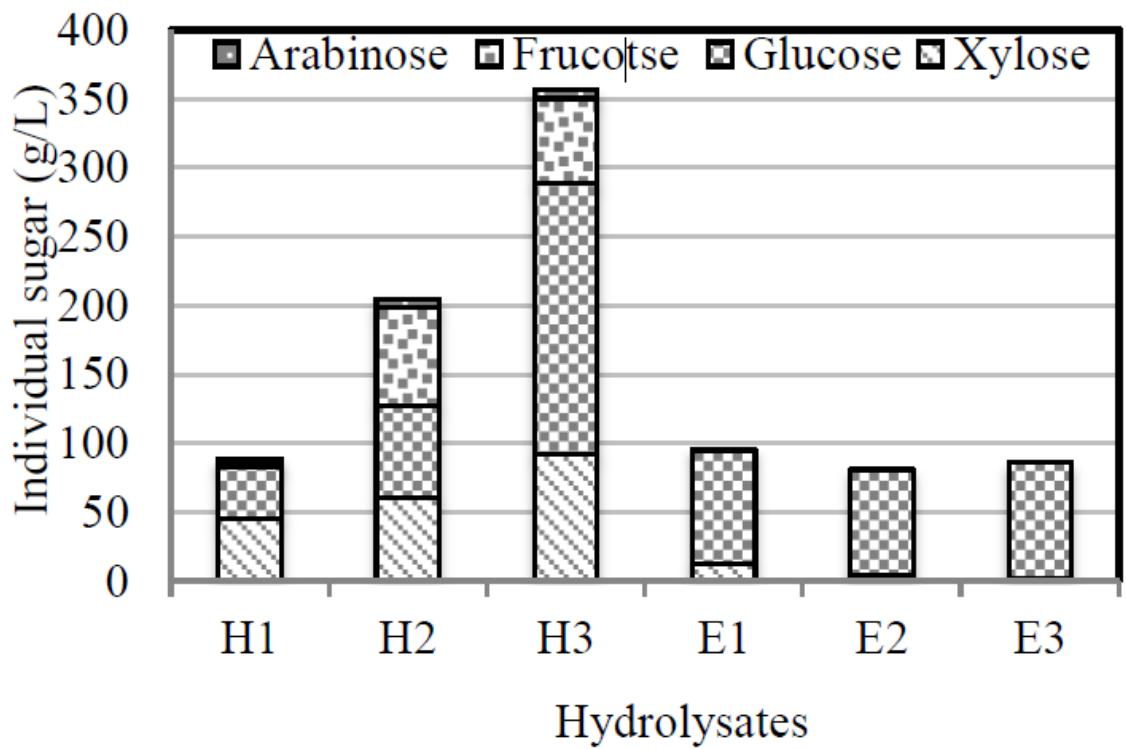

**Fig.1.** Individual sugar compositions of pineapple peel waste. H1, H2, H3: hydrolysate sugars of binary acids (sulfuric acid plus phenol) after pretreatment at different temperature 121, 150, 180 ºC, respectively. E1, E2, E3: individual sugar after enzymatic hydrolysis of residual biomass previously steam exploded for same pretreatment of hydrolysates

The phenol + sulfuric acid treatment resulted into considerable increase in total sugar release of 357.59 g/L with inhibitor formation (acetic acid 5.9 g/L, total phenolics 11.06 g/L, and total furan 6.35 mg/L) at 180 °C (Table 2). This enhancement in sugar release might be due to the presence of phenol which may act as free radical scavenger during hydrolysis. Lamminpää et al. (2015) has observed lower xylose dehydration into furfural when lignin (being phenolic acid) was added during sulfuric acid treatment. Further, increase in reaction temperature to 200 °C did not observe a significant improvement in sugar release (100-280 g/L) and lead to undesirable sugar degradations (Table 2). Ayeni et al. (2013) also showed decrease in cellulose recovery due to thermal degradation of cellulose treated at 195-200 °C. Besides, Pal et al. (2016a) also reported lower xylose yield of around 81% when operated at higher temperature up to 200 °C. Therefore, binary acid (phenol + sulfuric acid) at 180 °C was proved to be effective for improved sugar release from pineapple peel waste with lower fermentation inhibitor concentration. To achieve a competitive sugar yield by using single acid; relatively higher acid concentrations are desirable which further increases the inhibitor formation and affect overall process efficiency (Pal et al., 2016a).

3.2. Characterization of pretreated solid-fraction
3.2.1. Morphological analysis
The morphological changes of the untreated and pretreated pineapple peel were evaluated with the help of SEM analysis (Fig. 2). The SEM image of untreated sample showed intact surface with curve, uneven appearance, and well-arranged structure of cellulose, hemicellulose, and lignin (Fig. 2 A). Fig. 2 B shows the biomass surface after sulfuric acid treatment which was partially disintegrated suggesting that the sulfuric acid could hydrolyze hemicellulose leading to distorted and broken structure. Additionally, the surface is irregular with presence of some droplet or coalesced like structure.

The steam explosion treatment using binary acid (phenol + sulfuric acid) removed the amorphous cellulose and hemicellulose from inner part (Fig. 2 C). In addition, the lignin re deposition on surface of biomass can be seen in the form of corrugated surface which may affect the enzymatic hydrolysis. Khawas and Deka, (2016) performed the chemical treatment on banana peel and the microstructure was irregular with starch granules on its surface. The disintegrated structure of biomass is due to the hydrolysis of hemicellulose, and removal of lignin (Wang et al., 2015). Further, the re-deposition of lignin and droplet like structure on the surface of biomass may hinder cellulose hydrolysis (Pielhop et al., 2016).

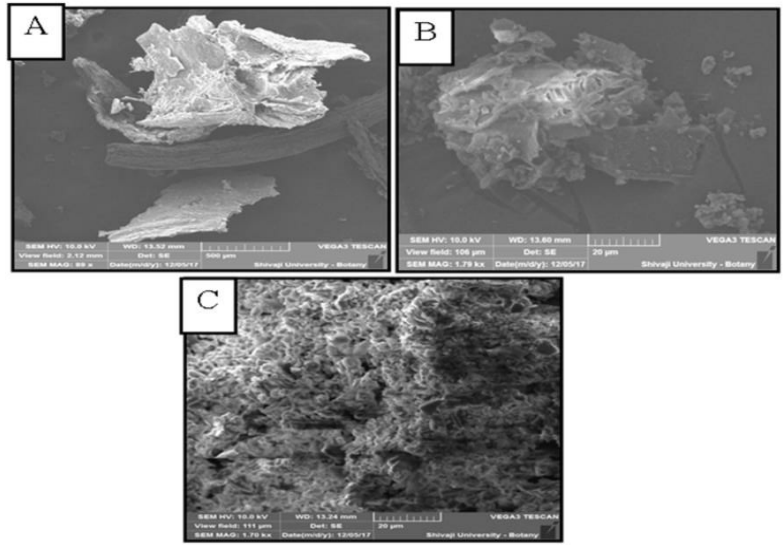

**Fig.2.** Scanning electron microscopy (SEM) images of pineapple peel A- untreated sample, B- treated using sulfuric acid at 180 °C, C- treated using binary acid (phenol plus sulfuric acid) at 180 °C

*3.2.2. Spectroscopy analysis - FTIR*

FTIR analysis was undertaken to reveal the changes in cellulose and hemicellulose structure of pineapple peel waste before and after pretreatment. It also gives information on the structure of lignin (syringol/guaiacol ratio) and hydrogen bonding alteration. The changes in the spectra were observed due to part of solid fractionated into hydrolysates during pretreatment. Fig. 3 shows FTIR spectra of untreated and pretreated pineapple peel. The variations in peak intensities for C-H, C-O, O-H, N-H, C-N bonds arose due to several pretreatment conditions. A band at 890 (1/cm) was associated with the characteristic absorption of b-(1-4) glycosidic bond of cellulose and is attributed to amorphous cellulose. As compared to untreated sample, the absorbance intensity after pretreatment was elevated which suggest the enhancement of cellulose content (Fig. 3 A and B). Li et al. (2016) detailed the steam explosion of *M. lutarioriparius* samples which showed increased peak intensity at 890 (1/cm) and also suggested the augmentation of cellulose content after treatment which is in-line with the current study.

The peak intensities of 1000-1200 (1/cm) can be endorsed to contributions of holocellulose having maxima at 1040 (1/cm), due to C-O stretching and 1165 (1/cm) due to the asymmetrical C-O-C stretching. The band absorption at 1247 (1/cm) arises due to C-O stretching and this absorption region indicate feature of hemicellulose as well as of lignin (Fig. 3 A and B). Lower band intensity at 1247 (1/cm) indicate removal of hemicellulose after pretreatments (Pal et al., 2016a). Surprisingly, steam exploded solid residue (at 180 °C) using binary acid (phenol plus sulfuric acid) showed no change in band intensity (at 1247 (1/cm)) which still contributes to higher sugar release of 357.59 g/L in hydrolysate. Incidentally, the band intensity for individual sulfuric acid treatment was dropped (0.368) which does not reflect to an increase in total sugar release (Fig. 3 C and D). This could be because of all hemicellulose solubilized does not get rehabilitated to xylose as a result of side reactions during pretreatment. However, detailed investigation on inhibitor minimization is essential in future studies that will open up a new era for second generation biochemical production.

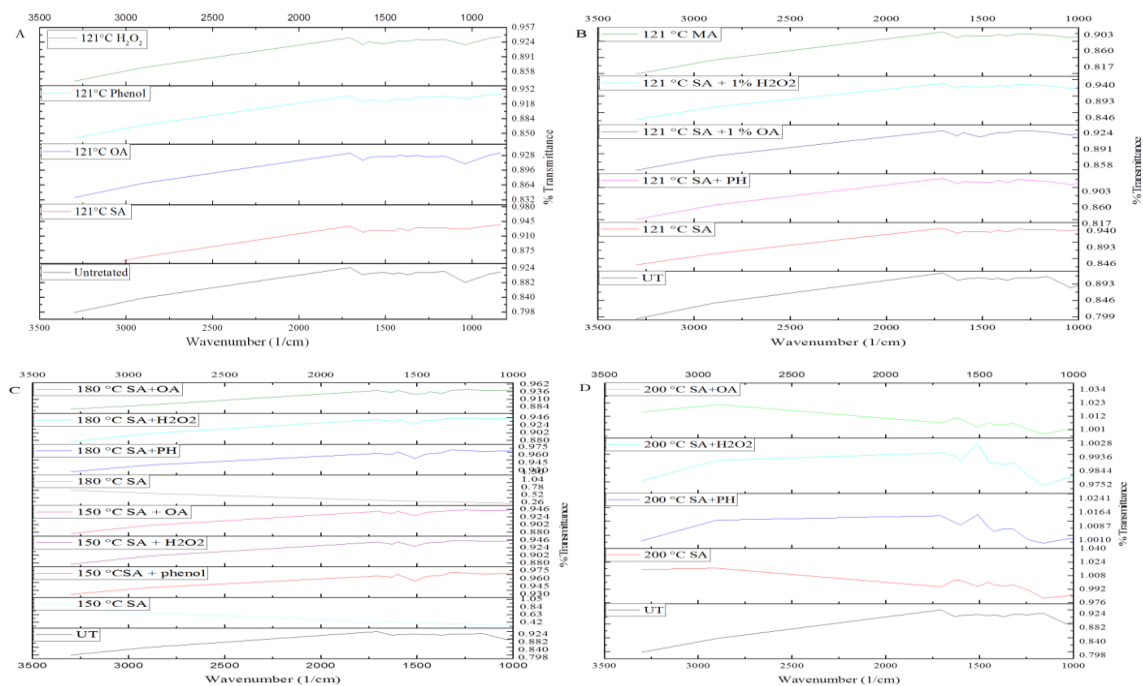

**Fig.3.** FTIR spectra of treated pineapple peel: A- Spectra of 121 ºC treated sample, B, C, D- spectra of treated sample at 150, 180, and 200 ºC, respectively. UT: Untreated sample; SA: Sulfuric acid treated; OA: Oxalic acid; PH: Phenol; H2O2: Hydrogen peroxide; SA+ OA: Combination of sulfuric acid with oxalic acid; SA+H2O2: Binary acid treatment with sulfuric acid and hydrogen peroxide; SA+PH: Treatment with sulfuric acid and phenol; MA: mixed acid pretreatment are the combinations of single acid enhancer

The crystallinity index is one of the important parameters to be considered during enzymatic hydrolysis. Hence, it was evaluated based on TCI, LOI, and HBI. Usually, elevated TCI and LOI values indicate highest degree of crystallinity and a more ordered cellulose structure. While lowest TCI and LOI values designate the amorphous structure of cellulose (Poletto et al., 2014). Table 3 summarizes the values of different crystallinity ratios (TCI, LOI, and HBI). Particularly, no significant difference in TCI ratio was observed between treated and untreated sample except for sulfuric acid treatment alone (at 150 and 180 °C). Like, LOI of untreated sample was merely same with one which is treated using binary acid (Table 3).

Further, the sample treated with single acid such as sulfuric acid, oxalic acid, phenol, hydrogen peroxide resulted into slightly lowered LOI which implies to the amorphous structure of cellulose. Interestingly, increased LOI values for sulfuric acid treatment at 150 and 180 °C were observed which might be due to more ordered structure of cellulose. Besides, HBI of pretreated sample was significantly higher than untreated sample. It may be due to cellulose chain is being in highly organized form and arranged in crystalline structure (Auxenfans et al., 2017).

FTIR also provides information on syringol (S) to guaiacol (G) ratio at wavenumber (1/cm) of 1260 and 1330 which represent the lignin in biomass. For untreated pineapple peel, S/G ratio was 0.89, a bit lower than that observed for different pretreatments (1.00). This increased pattern of S/G ratio is interrelated with the enzymatic hydrolysis. All the pretreatment studies showed an elevated S/G ratio which is expected to enhance enzymatic hydrolysis efficiencies. Pal et al., (2016a) and Guo et al. (2014) reported that an increase in S/G ratio enhanced the enzymatic hydrolysis efficiency.

**Table 3** Infrared crystallinity ratio of untreated and treated sample of pineapple peel at 121, 150 and 180 °C

**Pretreatment at 121 °C**

| Particulars | UT | 1% SA | 1% OA | 1% PH | 1% $H_2O_2$ | SA+ OA (1.3+1%) | SA+$H_2O_2$ (1.3+1%) | SA+PH (1.3+1%) | MA (1%) |
|---|---|---|---|---|---|---|---|---|---|
| Total crystallinity index | 1.08 | 1.08 | 1.06 | 1.06 | 1.06 | 1.05 | 1.06 | 1.07 | 1.07 |
| Lateral order index | 1.05 | 0.98 | 0.99 | 0.98 | 0.99 | 1.00 | 1.00 | 1.01 | 1.00 |
| Hydrogen bond intensity | 0.86 | 0.88 | 0.89 | 0.89 | 0.89 | 0.911 | 1.12 | 1.13 | 1.13 |
| Syringol to guaiacol ratio | 0.89 | 1.00 | 1.00 | 1.00 | 1.00 | 1.00 | 1.00 | 1.00 | 1.00 |

**Steam explosion at 150 and 180 °C**

| | UT | 150 SA | 150 SA+ OA | 150 SA+$H_2O_2$ | 150 SA+PH | 180 SA | 180 SA+OA | 180 SA+$H_2O_2$ | 180 SA+PH |
|---|---|---|---|---|---|---|---|---|---|
| Total crystallinity index | 1.08 | 0.387 | 1.00 | 1.04 | 1.02 | 0.57 | 1.02 | 1.04 | 1.02 |
| Lateral order index | 1.05 | 1.21 | 0.92 | 0.99 | 0.99 | 1.21 | 0.95 | 0.99 | 0.99 |
| Hydrogen bond intensity | 0.86 | 2.13 | 0.91 | 0.92 | 0.95 | 2.13 | 0.98 | 0.92 | 0.95 |
| Syringol to guaiacol ratio | 0.89 | 0.97 | 1.00 | 1.00 | 1.00 | 0.97 | 1.02 | 1.00 | 1.00 |

UT: Untreated sample; SA: Sulfuric acid treated; OA: Oxalic acid; PH: Phenol; $H_2O_2$: Hydrogen peroxide; SA+ OA: Combination of sulfuric acid with oxalic acid; SA+$H_2O_2$: Binary acid treatment with sulfuric acid and hydrogen peroxide; SA+PH: Treatment with sulfuric acid and phenol; MA: mixed acid pretreatment are the combinations of single acid enhancer

3.3. Enzymatic hydrolysis of pretreated solid-residue

During steam explosion; cellulose was significantly conserved in solid-fraction and its accessibility increased with elevated pretreatment temperatures (Ayeni and Daramola, 2017; Gaur et al., 2017). Therefore, solid-fractions obtained after pretreatment at 121 °C (using single acid, binary acid, and mixed acid) and steam explosion at 150 and 180 °C (using single acid and binary acid) were used for enzymatic hydrolysis experiments. The optimized enzyme operation conditions (10% w/v solid loading, 60 FPU enzyme/g substrate, pH 5.5, 50 °C, 120 rpm, and 72 h) were used from previous study (Harde et al., 2016). Tables 1 and 2 summarize the HPLC results of glucose release after enzymatic hydrolysis. The highest glucose release observed was in range of 92-113 g/L (pretreated at 121 °C using individual enhancer) (Table 1). On the other hand, cellulose conversion efficiency of samples (pretreated at 121 °C using binary acid) was found to be lowered and reason behind it is still unclear. Besides, the lower glucose release in the range of 80-90 g/L was obtained during enzymatic hydrolysis of steam exploded residue (150 and 180 °C). This may be due to the removal of substantial amount of amorphous cellulose in liquid hydrolysate during acid treatment itself (Fig. 1). Another reason could be re-deposition of lignin which can be seen in Fig. 2 C that may hinder enzymatic hydrolysis. Auxenfans et al. (2017) also highlighted the re-polymerized/re-accumulated lignin affects enzymatic hydrolysis leading to lower glucose release.

In case of steam exploded residue; elevated LOI and HBI levels (Table 3) indicated increased crystallinity with highly ordered cellulose structure which further decreased the enzymatic efficiency (Auxenfans et al., 2017; Ayeni and Daramola, 2017). Overall, the enzymatic efficiency was greater for residues treated at relatively lower temperature under varied enhancer and followed the order: phenol > sulfuric acid > oxalic acid > hydrogen peroxide >1% phenol + sulfuric acid >1% hydrogen peroxide + sulfuric acid >1% oxalic acid + sulfuric acid.

3.4. Mixed sugar (pentose and hexose) fermentation

The liquid-fraction obtained after phenol + sulfuric acid treatment at 180 °C was directly fermented to yield butanol via ABE fermentation using *C. acetobutylicum* NRRL B 527. Among different Clostridial spp., *C. acetobutylicum* is most advantageous, well studied and appropriate strain conventionally being used during biobutanol production from past many decade. Moreover, it consumes variety of sugars (hexose as well as pentose) which are beneficial during second generation biomass utilization (Bankar et al., 2013). The liquid-fraction obtained after pre-treatment (phenol + sulfuric acid treatment at 180 °C) without detoxification step was used during fermentation experiments that resulted in lower solvent production of 0.99 g/L with higher acid level upto 11.1 g/L. This may be due to the presence of fermentation inhibitors which led to acid crash. Therefore, inclusion of detoxification technique (Nimbalkar et al., 2017; Harde et al., 2016) is required for successful ABE production. Hence, the liquid fraction was detoxified using activated charcoal.

Detoxification step reduced the fermentation inhibitor concentrations up to 0.1 g/L total phenolics, 0.1 mg/L total furans, and 3.1 g/L acetic acid. The fermentation removal efficiency was >97% for total phenolics, ~98% for total furans, and 50% for acetic acid with acceptable total sugar loss (less than 12%). Furthermore, the detoxified slurry (liquid-fraction after steam explosion) and liquid fraction obtained after enzymatic hydrolysis were pooled together and subjected to batch fermentation till 120 h and samples were taken at an interval of 24 h for analysis.

The fermentation of pooled fraction resulted into 9.28 g/L total ABE and 2.6 g/L total acids with 61.66% sugar utilization. The sugar consumption profile was: glucose > fructose > xylose > arabinose. Solvent yield (0.25 g ABE/g sugar consumed) and productivity (0.077 g/(L.h)) obtained were very close to the control (P2) medium (0.26 g ABE/g sugar consumed and 0.13 g/(L.h)). The detailed ABE production profile using control and hydrolysate are shown in Fig. 4.

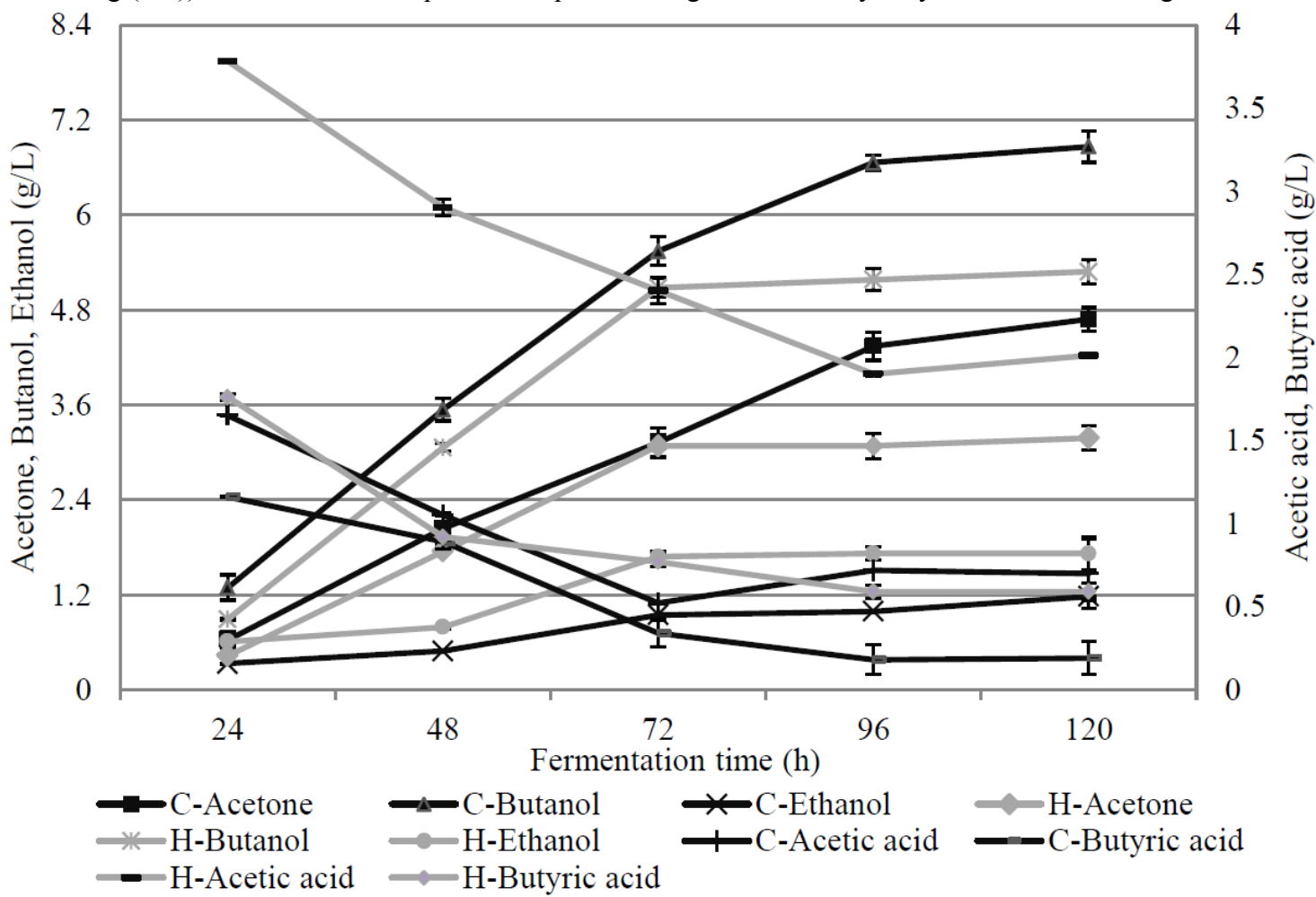

**Fig.4.** Fermentation profiles of butanol, acetone, and ethanol production by *Clostridium acetobutylicum* NRRL B 527; Solid dark color line indicate control (C); grey color line indicate hydrolysate (H)

An elevated acid level (2.6 g/L) in fermentation hydrolysate could be responsible for decreased total solvent production as it severely impedes Clostridial growth and solvent production. Incidentally, many researchers have produced biobutanol from varied feedstock by incorporating *C. acetobutylicum* as test organism. A comparative evaluation of literature reports with current study on ABE concentration, yield, and productivity using *C. acetobutylicum* are as shown in Table 4.

**Table 4** Comparison of ABE yield and productivity from current study and reported ABE yield from lignocellulosic biomass

| Feedstock | Microorganism | ABE (g/L) | ABE yield (g/g) | Productivity (g/(L.h)) | References |
|---|---|---|---|---|---|
| Pineapple peel waste | *C. acetobutylicum* NRRL B 527 | 9.08 | 0.25 | 0.096 | Current study |
| Norway spruce chips | *C. acetobutylicum* DSM 1731 | 10.6 | 0.35 | 0.11 | Yang et al. 2018 |
| Municipal solid waste | *C. acetobutylicum* NRRL B-591 | 6.2 | 0.21 | 0.06 | Farmanbordar et al. 2018 |
| Cauliflower waste | *C. acetobutylicum* NRRL B 527 | 5.35 | 0.17 | 0.05 | Khedkar et al. 2017b |
| Oil palm empty fruit bunch | *C. acetobutylicum* ATCC 824 | 4.45 | 0.18 | 0.06 | Ibrahim et al. 2015 |
| Roots of *Coleus forskohlii* | *C. acetobutylicum* NCIM 2877 | 5.32 | 0.20 | 0.05 | Harde et al. 2016 |
| Spent liquor | *C. acetobutylicum* DSM 792 | 8.79 | 0.20 | 0.09 | Survase et al. 2011 |
| Apple peels | *C. acetobutylicum* DSM 792 | 20.0 | 0.28 | 0.05 | Raganati et al. 2015 |

3.5. Mass balance of biobutanol production

The preliminary mass balance of biobutanol production from steam explosion treatment was calculated (Fig. 5). 1 kg pineapple peel after pretreatment produced 357.48 g of total sugars (xylose, glucose, fructose, and arabinose) in liquid-fraction. Remaining solid-fraction (539.79 g) produced 85.77 g sugars (glucose and xylose) after enzymatic hydrolysis. The pooled solid and liquid fractions generated 411.55 g fermentable sugars that produced 62.22 g ABE containing 35.5 g butanol.

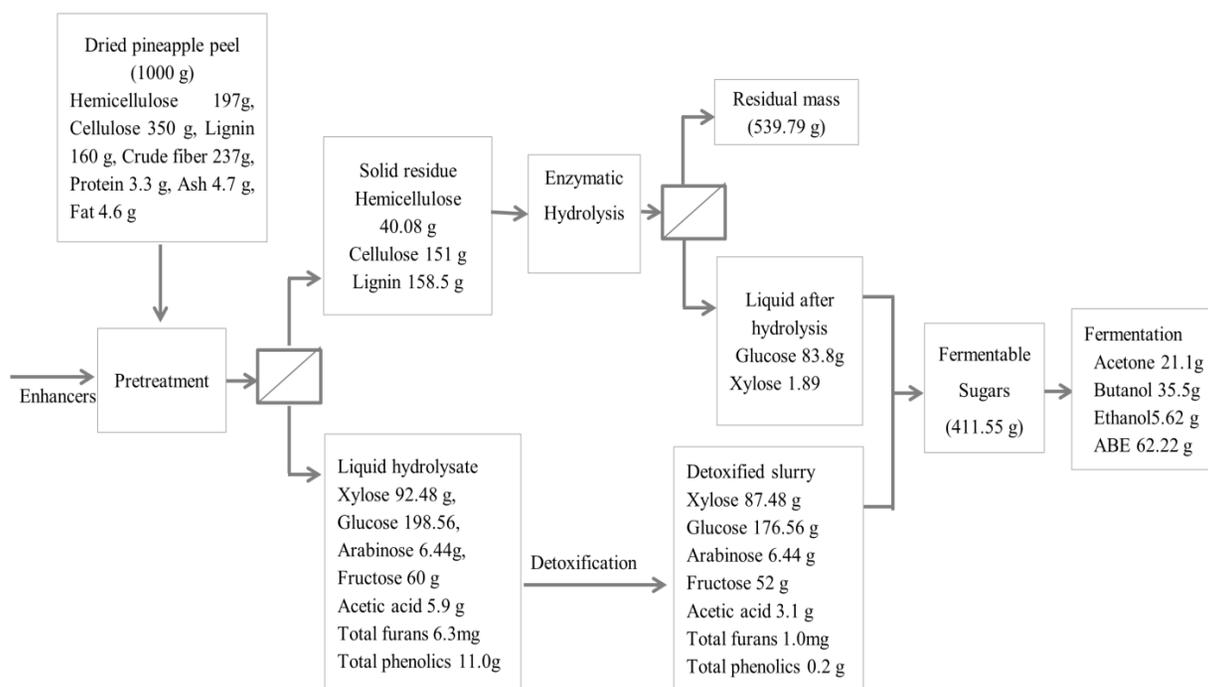

**Fig.5.** Mass balance for butanol production from pineapple peel waste pretreated using steam explosion and subsequent enzymatic hydrolysis and ABE fermentation

The fermentation experiments are not optimized in current study and final results will change drastically under optimized conditions. Hence, it would be appropriate to calculate detailed techno-economic and life cycle analysis including cost-profit assessment after the process is ready for its pilot scale demonstration. Farmanbordar et al. (2018) showed the mass balance for biobutanol production from biodegradable fraction of municipal solid waste (BMSW) and reported the butanol of 83.9 g, acetone 36.6 g, and ethanol 20.8 g from each kg of BMSW. Furthermore, the mass balance of butanol production from hydrogen peroxide-acetic acid pretreated spruce resulted in 575 g fermentable sugars which produced 201.2 g ABE (Yang et al., 2018).

From the biorefinery and economy point of view, pineapple peel hydrolysate is suggested as an appropriate substrate for ABE fermentation. Therefore, the binary acid enhancer and steam explosion process together demonstrates its feasibility at lab-scale. Furthermore, the efforts to scale up the steam explosion at pilot/industrial scale are still one of the interested areas to work with. Besides, various fermentation technological improvements such as fed-batch and continuous operation would further improve ABE solvent titer and yield to check its viability at large scale operation.

## 4. Conclusions

Screening of single acid pretreatment-enhancer and its binary combination provides insight on enhancement of monomeric sugar release and lowered inhibitor formation during pretreatment. The pretreatment at 180 °C using phenol + sulfuric acid yielded maximum fermentable sugar (357.25 g/L) in hydrolysate while enzymatic hydrolysis of residual biomass obtained 85.77 g/L hexose sugar. SEM and FTIR studies revealed the physiological changes and an elevated S/G ratio, respectively after treatment. Further, the fermentability was tested using *C. acetobutylicum* NRRL B 527 and resulted in highest biobutanol titer of 5.18 g/L with 2.6 g/L total acid using hydrolysate obtained at 180 °C treatment. The synergistic action of binary acid showed positive effect on detoxification as well as on biobutanol production.

**Acknowledgments**
The authors gratefully acknowledge Department of Science and Technology (DST) of Ministry of Science and Technology, Government of India, for providing financial support under the scheme of DST INSPIRE faculty award, (IFA 13-ENG-68/July 28, 2014) during the course of this investigation. Authors are thankful to Dr. Rahul Bhambure and Dr. Manoj Kamble from CSIR- National Chemical Laboratory, Pune for their valuable input during analysis. Authors are also thankful to Dr. Mansingraj Nimbalkar from Shivaji University, Kolhapur for SEM analysis.

**Appendix A. Supplementary data**
Supplementary data related to this article can be found at https://doi.org/10.1016/j.jclepro.2018.07.205.